\documentclass[camera]{jpaper}

\usepackage{multirow}
\usepackage{mathtools}
\usepackage{xspace}
\usepackage{textcomp}
\usepackage{listings}
\usepackage[plain]{algorithm}  
\usepackage[noend]{algpseudocode}
\usepackage{algorithmicx}
\floatname{algorithm}{Template}
\usepackage{xfrac}
\usepackage{subscript}

\usepackage{dblfloatfix}
\usepackage{appendix}

\usepackage{float}
\usepackage{booktabs}
\usepackage{placeins}
\usepackage{balance}
\usepackage[absolute]{textpos}
\usepackage[nocompress]{cite}
\usepackage{array}

\makeatletter
\let\MYcaption\@makecaption
\makeatother

\usepackage[font=footnotesize]{subcaption}

\makeatletter
\let\@makecaption\MYcaption
\makeatother

\usepackage{fixltx2e}
\usepackage{dblfloatfix}
\usepackage[nolessnomore, italic]{mathastext}
\usepackage[T1]{fontenc}
\usepackage[usenames,dvipsnames,svgnames,table]{xcolor}
\usepackage[normalem]{ulem}
\usepackage{enumitem}
\usepackage{setspace}
\usepackage{indentfirst}
\usepackage{footmisc}
\usepackage{fancyhdr}
\usepackage{authblk}
\usepackage[us,12hr]{datetime}
\usepackage[keeplastbox]{flushend}
\usepackage[hidelinks]{hyperref}

\newif\ifcameraready
\camerareadytrue

\newcommand{\versionnum}[0]{5}

\ifcameraready
\else
\fi

\ifcameraready
  \newcommand{\todo}[1][]{}
  \newcommand{\ch}[0]{}
\else
  \newcommand{\todo}[1][]{\textbf{\fcolorbox{black}{red}{\color{white}{TODO}}} \underline{$\overline{\hbox{\emph{#1}}}$}}
  \newcommand{\ch}[1]{{\color{BrickRed} #1}}
\fi

\usepackage{ifthen}
\usepackage{calc}
\usepackage{pifont}
\usepackage{color}
\usepackage{fancyhdr}
\usepackage{tikz}
\usepackage{anyfontsize}
\usepackage{gensymb}

\newcounter{hours}
\newcounter{minutes}

\newcommand{\myitem}[1]{\emph{(#1)}\xspace}

\newcommand{\varr}[0]{{{$V_{array}$}}\xspace}

\newcommand{\vmin}[0]{$V_{min}$\xspace}

\newcommand{\squishlist} {
    \begin{list}{$\bullet$} {
        \setlength{\itemsep}{-2pt}
            \setlength{\parsep}{2pt}
            \setlength{\topsep}{0pt}
            \setlength{\partopsep}{0pt}
            \setlength{\leftmargin}{1.0em}
            \setlength{\labelwidth}{1em}
            \setlength{\labelsep}{0.5em}
    }
}
\newcommand{\squishend} {
    \end{list}
}

\newboolean{timeofmake}
\setboolean{timeofmake}{false}

\newcommand{\ignore}[1]{}

\newcommand{\highlight}[1]{#1}
\newcommand{\changes}[1]{#1}

\newcommand{\new}[1]{#1}
\newcommand{\fix}[1]{#1}
\newcommand{\fixII}[1]{#1}
\newcommand{\fixIII}[1]{#1}
\newcommand{\fixIV}[1]{#1}
\newcommand{\fixV}[1]{#1}
\newcommand{\response}[1]{#1}

\newcommand{\voltron}[0]{{Voltron}\xspace}
\newcommand{\memdvfs}[0]{{MemDVFS}\xspace}

\newcommand{\tras}[0]{{{tRAS}}\xspace}
\newcommand{\trp}[0]{{tRP}\xspace}

\newcommand{\trcd}[0]{{{tRCD}}\xspace}

\newcommand{\act}[0]{\texttt{\small{ACTIVATE}}\xspace}

\newcommand{\figputHW}[2]{
\begin{figure}[h]
\begin{minipage}{\linewidth}
\footnotesize 
\begin{center}
\includegraphics[width=1.0\linewidth]{plots/#1}
\end{center}
\vspace{-0.2in}
\caption{#2 \label{fig:#1}}
\end{minipage}
\end{figure}
}

\newcommand{\figputHS}[3]{
\begin{figure}[h]
\begin{minipage}{\linewidth}
\begin{center}
\includegraphics[scale=#2]{plots/#1}
\end{center}
\vspace{-0.1in}
\caption{#3 \label{fig:#1}}
\end{minipage}
\end{figure}
}

\newcommand{\figref}[1]{Figure~\ref{fig:#1}}
\newcommand{\tabref}[1]{Table~\ref{tab:#1}}

\newcommand{\secref}[1]{Section~\ref{sec:#1}}
\newcommand{\ssecref}[1]{Section~\ref{ssec:#1}}

\newcommand{\paratitle}[1]{\textbf{#1.}\xspace}

\title{Voltron: Understanding and Exploiting\\ the Voltage--Latency--Reliability
  Trade-Offs\\ in Modern DRAM Chips to Improve Energy Efficiency}

\author{%
{Kevin K. Chang$^{1,2}$}%
\qquad%
{Abdullah Giray Ya\u{g}l{\i}k\c{c}{\i}$^{2}$}%
\qquad%
{Saugata Ghose$^{2}$}%
\qquad%
{Aditya Agrawal$^{3}$}%
\vspace{2pt}\\
{Niladrish Chatterjee$^{3}$}%
\qquad%
{Abhijith Kashyap$^{4,2}$}%
\qquad%
{Donghyuk Lee$^{3}$}%
\vspace{2pt}\\
{Mike O'Connor$^{3,5}$}%
\qquad%
{Hasan Hassan$^{6}$}%
\qquad%
{Onur Mutlu$^{6}$}%
}
\affil{%
{\it%
$^1$Facebook%
\qquad%
$^2$Carnegie Mellon University%
\qquad%
$^3$NVIDIA Research%
}\vspace{2pt}\\{\it
$^4$NVIDIA%
\qquad%
$^5$The University of Texas at Austin%
\qquad%
$^6$ETH Z{\"u}rich%
}}

\date{}

\begin{document} \sloppy 
\maketitle

\begin{abstract}

\new{This paper summarizes our work on experimental characterization and
  analysis of reduced-voltage operation in modern DRAM chips, which was
  published in SIGMETRICS 2017~\cite{chang-sigmetrics2017}, and examines the work's
  significance and future potential. This work is motivated
  to reduce the energy consumption of DRAM, which} is a critical concern in
modern computing systems. Improvements in manufacturing process technology have
allowed DRAM vendors to lower the DRAM supply voltage conservatively, which
reduces some of the DRAM energy consumption. We would like to reduce the DRAM
supply voltage more aggressively, to further reduce energy. Aggressive supply
voltage reduction requires a thorough understanding of the effect voltage
scaling has on DRAM access latency and DRAM reliability.

We take a comprehensive approach to understanding and exploiting the latency and
reliability characteristics of modern DRAM when the supply voltage is lowered
below the \fixII{nominal voltage level} specified by \fixIV{DRAM standards}.
Using an \mbox{\new{open-source}} FPGA-based testing platform \new{based on
  SoftMC~\cite{hassan-hpca2017}}, we perform an experimental study of 124 real
DDR3L (low-voltage) DRAM chips manufactured recently by three major DRAM
vendors. We find that reducing the supply voltage below a certain point
introduces bit errors in the data, and we comprehensively characterize the
behavior of these errors. We discover that these errors can be avoided by
increasing the latency of three major DRAM operations (activation, restoration,
and precharge). \fixII{We perform} detailed DRAM circuit simulations to validate
and explain our experimental findings. We also characterize the various
relationships between reduced supply voltage and error locations, stored data
patterns, DRAM temperature, and data retention.

Based on our observations, we propose a new DRAM energy reduction mechanism,
called \emph{Voltron}. The key idea of Voltron is to use a performance model to
determine \fix{by} how much we can reduce the supply voltage without introducing errors
and without exceeding a user-specified threshold for performance loss. Our
evaluations show that Voltron reduces the average DRAM and system energy
consumption by 10.5\% and 7.3\%, respectively, while limiting the average system
performance loss to only 1.8\%, for a variety of \fix{memory-intensive} quad-core
workloads. We also show that Voltron significantly outperforms prior dynamic
voltage and frequency scaling mechanisms for DRAM. \new{We believe our
  experimental characterization and findings can pave the way for new mechanisms
that exploit DRAM voltage to improve power, performance, energy, and
reliability.}

\end{abstract}

\section{Motivation}

In a wide range of modern computing systems, spanning from warehouse-scale data
centers to mobile platforms, energy consumption is a first-order
concern\new{~\cite{hoelzle-book2009,ibm-power7,deng-asplos2011,mutlu-imw2013,mudge-computer2001}}.
In these systems, the energy consumed by the DRAM-based main memory system
constitutes a significant fraction of the total energy.  For example,
\fix{experimental} studies of production systems have shown that DRAM consumes
40\% of the total energy in servers~\cite{hoelzle-book2009,ibmpower7-hpca}
\fixII{and 40\%} of the total power in graphics cards~\cite{paul-isca2015}.

Improvements in manufacturing process technology have allowed DRAM vendors to
lower the DRAM supply voltage conservatively, which reduces some of the DRAM
energy consumption~\cite{jedec-ddr3l, jedec-lpddr3, jedec-lpddr4}. In this work, we would like
to reduce DRAM energy by \emph{further reducing DRAM supply voltage}. Vendors
choose a conservatively high supply voltage, to provide a \emph{guardband} that
allows DRAM chips with worst-case process variation to operate without errors
\fixIV{under the worst-case operating conditions}~\cite{david-icac2011}. The
exact amount of supply voltage guardband varies across chips, and lowering the
voltage below the guardband can result in erroneous or even undefined
behavior~\cite{chang-sigmetrics2017}.
Therefore, we need to understand how DRAM chips behave during reduced-voltage
operation. To our knowledge, no previously published work examines the effect of
using a wide range of different supply voltage values on the reliability,
latency, and retention characteristics of DRAM chips.


\changes{\textbf{Our goal} in our \new{SIGMETRICS 2017} paper~\cite{chang-sigmetrics2017}
is to \myitem{i}~characterize and understand the relationship between supply
voltage reduction and various characteristics of DRAM, including DRAM
reliability, latency, and data retention; and \myitem{ii}~use the insights
derived from this characterization and understanding to design a new mechanism
that can aggressively lower the supply voltage to reduce DRAM energy consumption
while keeping performance loss under a bound.}


To this end, we build an FPGA-based testing platform based on
SoftMC~\cite{hassan-hpca2017} 
that allows us to tune the DRAM
supply voltage and change DRAM timing parameters (i.e., the amount of time the
memory controller waits for a DRAM operation to complete). We perform an
experimental study on 124 real 4Gb DDR3L (low-voltage) DRAM chips manufactured
recently (between 2014 and 2016) by three major DRAM vendors. Our extensive
experimental characterization yields four major observations on how DRAM
\highlight{latency, reliability, and data retention are} affected by reduced
voltage.

Based on our experimental observations, we propose a new low-cost DRAM energy
\fix{reduction} mechanism called \emph{Voltron}. The key idea of
  Voltron is to use a performance model to determine \fixIV{by} how much we can
  reduce the DRAM array voltage at runtime without introducing errors and
  without exceeding a user-specified threshold for \fixII{acceptable}
  performance loss.


\section{Characterization of DRAM Under \\ Reduced Supply Voltage}
\label{sec:results}

In this section, we briefly summarize our four major observations from our
detailed experimental characterization of 31 commodity DRAM modules, also called
DIMMs, from three vendors, when \changes{the DIMMs} operate under \fix{reduced}
supply voltage (i.e., below the nominal voltage level of 1.35V). Each DIMM
comprises 4 DDR3L DRAM chips, totaling to 124 chips for 31 DIMMs. Each chip has
a 4Gb density. Thus, \fixII{each} of our DIMMs \fixII{has} a 2GB capacity.
\tabref{dimm_list} describes the relevant information about \fixIV{the} tested
DIMMs. For a complete discussion on all of our observations and experimental
methodology, we refer the reader to our \new{SIGMETRICS 2017}
paper~\cite{chang-sigmetrics2017}.

\begin{table}[h]
  \small
  \centering
    \setlength{\tabcolsep}{.35em}
    \begin{tabular}{cccc}
        \toprule
        \multirow{2}{*}{Vendor} & Total Number & Timing (ns) &
        Assembly  \\
        & of Chips & (\trcd/\trp/\tras) & Year  \\
        \midrule
        A (10 DIMMs) & 40 & 13.75/13.75/35 & 2015-16\\
        B (12 DIMMs) & 48 & 13.75/13.75/35 & 2014-15 \\
        C (9 DIMMs) & 36 & 13.75/13.75/35 & 2015 \\
        \bottomrule
    \end{tabular}
  \caption{\fixIII{Main} properties of the tested DIMMs. Reproduced from
    \cite{chang-sigmetrics2017}.}
  \label{tab:dimm_list}
\end{table}

\subsection{DRAM Reliability as Voltage Decreases}

We first study the reliability of DRAM chips under low voltage, which was not
studied by prior works on DRAM voltage scaling
\fix{(e.g.,~\cite{david-icac2011}; see Section 4 for a detailed discussion of
  these works).} \figref{dimm_errors_all} shows the fraction of cache lines that
\changes{experience} at least 1~bit of error (i.e., \emph{1~bit flip}) in each
DIMM \changes{(represented by each curve)}, \fixIV{categorized based on vendor.}

\figputHW{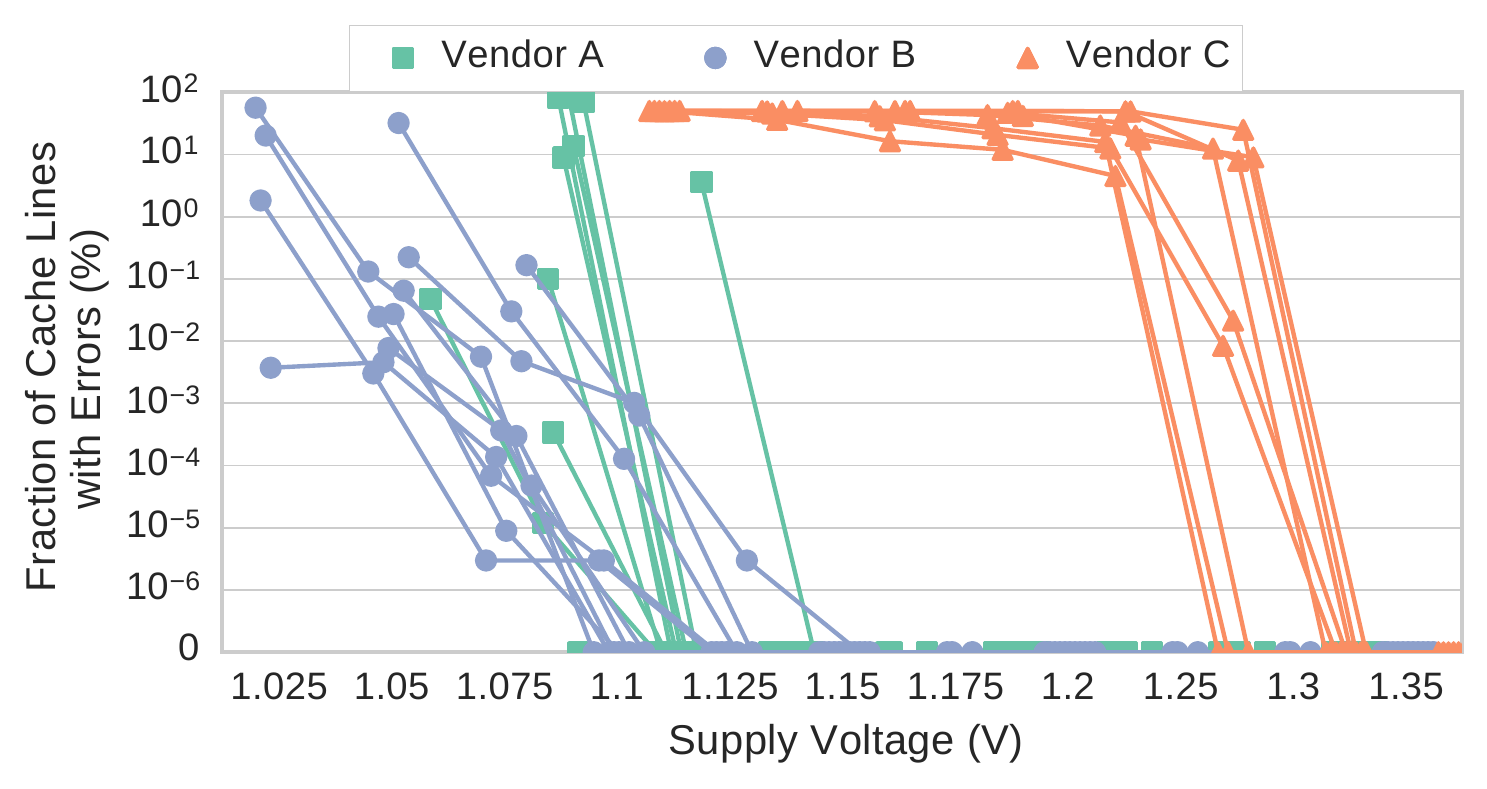}{The fraction of erroneous cache lines in each DIMM as
we reduce the supply voltage, with a fixed latency. Reproduced from
\cite{chang-sigmetrics2017}.}

We observe that we can reliably access data when DRAM supply voltage is lowered
below the nominal voltage level, {\em until a certain voltage value}, \vmin,
which is the minimum voltage level at which no bit errors occur. Furthermore, we
find that we can reduce the voltage below \vmin to attain further energy
savings, but that errors start occurring in some of the data read from memory.
However, not all cache lines exhibit errors for all supply voltage values below
\vmin. Instead, the number of erroneous cache lines \changes{for each DIMM}
increases as we reduce the voltage further below \vmin. Specifically, Vendor A's
DIMMs \fix{experience a} near-exponential increase \fix{in} errors as \fix{the}
supply voltage reduces below \vmin. This is mainly due to the
\new{\emph{manufacturing process}~\cite{lee-hpca2015} and \emph{architectural
  variation}~\cite{lee-sigmetrics2017}}, which introduces strength and size
variation across the different DRAM cells within a chip.

\new{We make two major conclusions: \myitem{i} the variation of errors due
  to reduced-voltage operation across vendors is very significant; and
  \myitem{ii} in most cases, there is a significant margin in the voltage
  specification, i.e., \vmin for each chip is significantly lower than the
  manufacturer-specified supply voltage value.}

\subsection{Longer Access Latency Mitigates \\ Voltage-Induced Errors}
\label{ssec:long_latency}

We observe that while reducing the voltage below \highlight{\vmin introduces bit
errors} in the data, we can prevent these errors if we increase the timing
parameters of three major DRAM operations, i.e., activation, restoration, and
precharge\new{~\cite{hassan-hpca2016,chang-sigmetrics2016,chang-sigmetrics2017,lee-hpca2015,lee-sigmetrics2017}}.\footnote{\new{We refer the reader to our prior
 works~\cite{chang-sigmetrics2016, kim-isca2012, lee-hpca2013,
      lee-hpca2015, kim-micro2010, kim-hpca2010,chang-hpca2016, hassan-hpca2016,
      chang-sigmetrics2017, lee-sigmetrics2017, lee-taco2016, lee-pact2015,
      liu-isca2012, liu-isca2013, patel-isca2017, chang-hpca2014,
      seshadri-micro2013, seshadri-micro2017, hassan-hpca2017, kim-cal2015,
      kim-isca2014, kim-hpca2018} for a detailed background on
  DRAM.}}
When the supply voltage is reduced, the DRAM cell capacitor charge takes a
longer time to change, thereby causing these DRAM operations to become slower to
complete. Errors are introduced into the data when the memory controller does
\emph{not} account for this \highlight{slowdown in the DRAM operations}.  We
find that if the memory controller allocates extra time for these operations to
finish when the supply voltage is below \vmin, errors no longer occur. We
\highlight{validate, analyze,} and explain this behavior using SPICE simulation
of a detailed circuit-level model, which we have \ch{openly} released
online~\cite{safari-github}. \new{Sections~4.1 and 4.2 of our SIGMETRICS 2017
  paper~\cite{chang-sigmetrics2017} provide our extensive circuit-level analyses,
  validated using data from real DRAM chips. }

\subsection{Spatial Locality of Errors}
\label{ssec:locality}

While reducing the supply voltage induces errors when the DRAM latency is \emph{not}
long enough, we also show that \new{\emph{not}} all DRAM locations experience errors at all
supply voltage levels. To understand the locality of the errors induced by a low
supply voltage, we show the probability of each DRAM row in a DIMM
\fix{experiencing} at least one bit of error across all experiments.

\begin{figure}[!h]
    \vspace{0.15in}
    \centering
    \subcaptionbox{DIMM B$_6$ of vendor~B at 1.05V.}[\linewidth][c]
    {
        \includegraphics[scale=1]{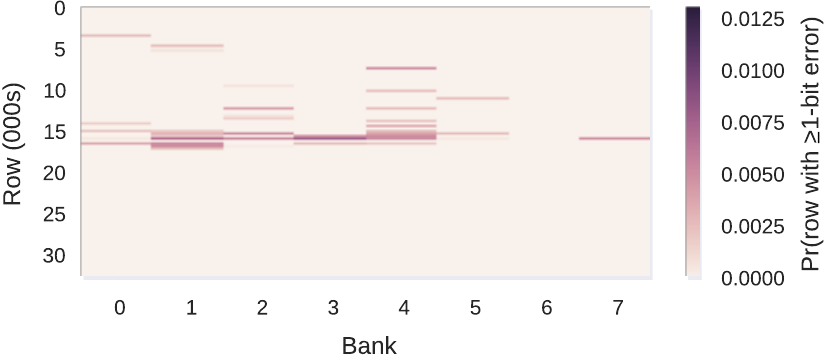}
    }
    \vspace{0.15in}

    \subcaptionbox{DIMM C$_2$ of vendor~C at 1.20V.}[\linewidth][c]
    {
        \includegraphics[scale=1]{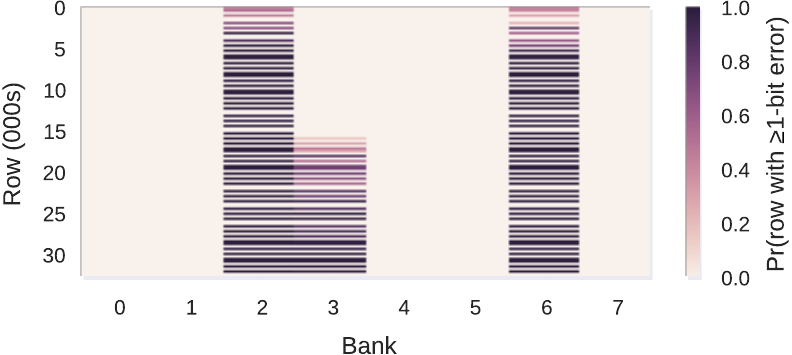}
    }
    \caption{The probability of error occurrence for two representative
      DIMMs, \fixIV{categorized into different rows and banks}, due to reduced voltage. Reproduced from
\cite{chang-sigmetrics2017}.}

    \label{fig:locality_C}
\end{figure}

\figref{locality_C} shows the probability of each row \fix{experiencing} at
least a one-bit error due to reduced voltage in the two representative DIMMs.
For each DIMM, we choose the supply voltage \new{at which} errors start appearing (i.e.,
the voltage \fix{level} one step below \vmin), and we do \emph{not} increase the
DRAM access latency \fixIV{(i.e., \new{keep it at} 10ns for both \trcd and \trp,
  \new{which are the activation and precharge timing parameters, respectively})}. The x-axis
and y-axis indicate the bank number and row number (in thousands), respectively.
Our tested DIMMs are divided into eight banks, and each bank consists of
32K~rows of cells. Additional results showing the error locations at different voltage
\fix{levels} are in our \new{SIGMETRICS 2017} paper~\cite{chang-sigmetrics2017}.


The major observation is that when only a small number of errors occur due to
reduced supply voltage, these errors tend to \emph{cluster} physically in
certain \emph{regions} of a DRAM chip, as opposed to being randomly distributed
throughout the chip.\footnote{\new{We believe this observation is due to both
    process and architectural variation across different regions in the DRAM
    chip.}} This observation implies that when we reduce the supply voltage to
the DRAM array, we need to increase the fundamental operation latencies for
\emph{only} the regions where errors can occur.

\subsection{Impact on Refresh Rate}

Commodity DRAM chips guarantee that all cells can safely retain data for 64ms,
after which the cells are \emph{refreshed} to replenish charge that leaks out of
the capacitors\new{~\cite{liu-isca2012,liu-isca2013,chang-hpca2014}}. We observe that
the effect of \fixIV{the supply voltage} on retention times is \emph{not}
statistically significant. Even when we reduce the supply voltage from 1.35V to
1.15V (i.e., a 15\% reduction), the rate at which charge leaks from the
capacitors is so slow that no data is lost during the 64ms refresh interval at
both 20$\celsius$ and 70$\celsius$. Therefore, we conclude that using a reduced
supply voltage does not require any changes to the standard refresh interval at
20$\celsius$ and 70$\celsius$. \new{Detailed results are in Section 4.6 of our
  SIGMETRICS 2017 paper~\cite{chang-sigmetrics2017}.}

%


\subsection{Other Experimental Observations}

We refer the reader to our SIGMETRICS 2017 paper~\cite{chang-sigmetrics2017} for
more details on the other two key observations. First, \new{we find that} the
most commonly-used ECC scheme, SECDED~\cite{luo-dsn2014, kang14,
  sridharan-asplos2015}, is unlikely to alleviate errors induced by a low supply
voltage. \new{This is because lowering voltage increases the fraction of data that
  contains more than two bits of errors, exceeding the one-bit correction
  capability of SECDED (see Section 4.4 of our SIGMETRICS 2017 paper~\cite{chang-sigmetrics2017})}. Second,
temperature affects the reliable access latency at low supply voltage levels and
the effect is very vendor-dependent (see Section 4.5 of our SIGMETRICS 2017 paper~\cite{chang-sigmetrics2017}).
\new{Out of the three major vendors whose DIMMs we evaluate, DIMMs from 
two vendors require \ch{longer activation and
  precharge latencies} to operate reliably at high temperature under low supply
  voltage. The main reason is that DRAM chips become slower at higher
  temperature~\cite{lee-sigmetrics2017, lee-hpca2015,chandrasekar-date2014}.}

\section{Exploiting Reduced-Voltage Behavior}
\label{sec:voltron}

Based on the extensive understanding we have developed on reduced-voltage operation
of real DRAM chips, we propose a new mechanism called \emph{\voltron}, which
reduces DRAM energy without sacrificing memory throughput. \voltron exploits the
fundamental observation that reducing the supply voltage to DRAM requires
increasing the latency of the three DRAM operations in order to prevent errors.
Using this observation, the key idea of Voltron is to use a performance model to
determine \fix{by how much to reduce the DRAM supply voltage}, without
introducing errors and without exceeding a user-specified threshold for
performance loss. \voltron consists of two main components: \myitem{i}
\emph{array voltage scaling} and \myitem{ii} \emph{performance-aware voltage
  control}.


\subsection{Components of Voltron}
\label{sec:voltron:components}

\noindent\textbf{Array Voltage Scaling.} Unlike prior works, Voltron does
\emph{not} reduce the voltage of the \emph{peripheral circuitry}, which is
responsible for transferring commands and data between the memory controller and
the DRAM chip. If Voltron were to reduce the voltage of the peripheral
circuitry, we would \emph{have to} also reduce the operating frequency of DRAM.
A reduction in the operating frequency reduces the memory data throughput, which
can significantly degrade the performance of applications that require high
memory bandwidth. Instead, Voltron reduces the voltage supplied to \emph{only}
the DRAM array without changing the voltage supplied to the peripheral
circuitry, thereby allowing the DRAM channel to maintain a high frequency while
reducing the power consumption of the DRAM array. To prevent errors from
occurring during reduced-voltage operation, Voltron increases the latency of the
three DRAM operations (activation, restoration, and precharge) based our
observation in \ssecref{long_latency}.

\new{

\noindent\textbf{Performance-Aware Voltage Control.} Array voltage scaling
provides system users with the ability to decrease DRAM array voltage (\varr) to
reduce DRAM power. Employing a lower \varr provides greater power savings, but
at the cost of longer DRAM access latency, which leads to larger performance
degradation. This trade-off varies widely across different applications, as each
application has a different tolerance to the increased memory latency. This
raises the question of how to pick a ``suitable'' array voltage level for
different applications as a system user or designer. For our evaluations, we say that
an array voltage level is suitable if it does not degrade system performance by
more than a user-specified threshold. Our goal is to provide a simple
\fixIII{technique} that can automatically select \fix{a} suitable \varr
\fix{value} for different applications. \changes{To this end, we propose
  \emph{performance-aware voltage control}, a \fix{power--performance} management
  policy that selects a minimum \varr which satisfies a desired performance
  constraint. The key observation is that an application's performance loss (due
  to increased memory latency) scales linearly with the application's memory
  \fix{demand \fixIII{(e.g., memory intensity)}.} Based on this
  \fixIV{empirical} observation \fixIV{we make}, we build a \emph{performance
    loss predictor} that leverages a linear model to predict an application's
  performance loss based on its characteristics \ch{and the effect of different
  voltage level choices} at runtime. Using the
  performance loss predictor, Voltron finds a value of \varr that \fix{can keep the}
  predicted performance within the user-specified target at runtime.} We refer the
reader to Section 5.2 of our SIGMETRICS 2017 paper~\cite{chang-sigmetrics2017} for
more detail and for an evaluation of the performance model \ch{alone}. }


\subsection{Evaluation}
\label{sec:voltron:eval}

We evaluate the system-level energy \ch{and performance} impact of \voltron using
Ramulator~\cite{kim-cal2015, safari-github}, integrated with
McPAT~\cite{mcpat:micro} and DRAMPower~\cite{drampower} for modeling the energy
consumption of both the processor and DRAM. Our workloads consist of 27
benchmarks from SPEC CPU2006~\cite{spec2006} and YCSB~\cite{cooper-socc2010}.
\new{We evaluate Voltron with a target performance loss of 5\%. \voltron
  executes the performance-aware voltage control mechanism once every four
  million cycles.} We refer the reader to Section~6.1 of our \new{SIGMETRICS
  2017 paper~\cite{chang-sigmetrics2017} for more detail on the} system configuration and workloads. We
qualitatively and quantitatively compare Voltron \fixIV{to} \textit{\memdvfs}, a
dynamic DRAM frequency and voltage scaling mechanism proposed by prior
work~\cite{david-icac2011}.

\figref{voltron_perf_e} shows the system energy savings \new{and the system
performance (i.e., weighted
speedup~\cite{eyerman-ieeemicro2008,snavely-asplos2000}) loss} due to
\mbox{\memdvfs} and \voltron, compared to a baseline DRAM with \fix{a supply
  voltage of} 1.35V. The graph uses box plots to show the distribution among all
workloads that are categorized as \fixV{either non-memory-intensive or
  memory-intensive}. The memory intensity is determined based on the
commonly-used metric MPKI (last-level cache misses per kilo-instruction). We
categorize an application as memory intensive when its MPKI is greater than or
equal to 15. We make two observations.

\figputHS{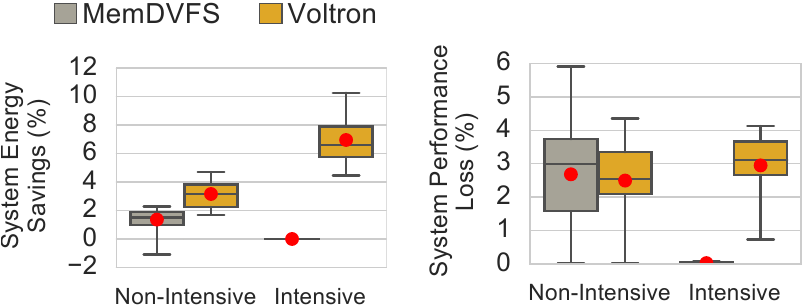}{1}{\new{Energy (left) and performance (right)} comparison between Voltron
and MemDVFS on non-memory-intensive and memory-intensive workloads. Adapted from
\cite{chang-sigmetrics2017}.}

\new{First, Voltron is effective and saves more energy than MemDVFS.} \memdvfs has
almost zero effect on \fixIV{memory-intensive} workloads. This is because
\memdvfs avoids scaling DRAM frequency \fix{(and hence voltage)} when an
application's memory bandwidth \fixIV{utilization} is above a fixed threshold.
Reducing the frequency can result in a large performance loss since the
\fixIV{memory-intensive} workloads require high memory throughput. As
\fix{memory-intensive} applications have high memory bandwidth consumption that
easily exceeds the \fixIV{fixed} threshold \fixIV{used by \memdvfs}, \memdvfs
\emph{cannot} perform frequency and voltage scaling during most of the execution
time. In contrast, \voltron reduces system energy by 7.0\% on average for
\fix{memory-intensive} workloads. \fix{Thus, we} demonstrate that \voltron is an
effective mechanism that improves \fixIV{system} energy efficiency not only on
\fix{non-memory-intensive} applications, but also \fixIV{(especially)} on
\fix{memory-intensive} workloads where prior work was unable to do so.

\new{\fix{Second, as shown in \figref{voltron_perf_e} (right)}, \voltron consistently
selects a \varr value that satisfies the performance loss bound of 5\% across
all workloads. \voltron incurs \fix{ an average (maximum) performance loss of
  2.5\% (4.4\%) and 2.9\% (4.1\%)} for \fixV{non-memory-intensive and
  memory-intensive} workloads, respectively. This demonstrates that our
performance model enables \voltron to select a low voltage value that saves
energy while bounding performance loss based on the user's requirement.

Our SIGMETRICS 2017 paper contains extensive performance and energy analysis of
the Voltron mechanism in Sections 6.2 to 6.8~\cite{chang-sigmetrics2017}. In
particular, we show that if we exploit spatial locality of errors
(\ssecref{locality}), we can improve the performance benefits of Voltron,
\ch{reducing the average performance loss for memory-intensive workloads
to 1.8\%}
(\ch{see} Section 6.5 of our SIGMETRICS 2017 paper~\cite{chang-sigmetrics2017}). We refer the reader to these sections for
a detailed evaluation of Voltron.
}



\section{Related Work}

\response{To our knowledge, this is the \fixIII{first work}} to \myitem{i}
experimentally \changes{characterize the reliability and performance of modern
  low-power DRAM chips under different supply voltages}, and \myitem{ii}
\fixIV{introduce} a new mechanism that reduces DRAM energy while retaining high
memory \fixIV{data} throughput by \fixIV{adjusting} the DRAM array voltage. We
briefly discuss other prior work in DRAM energy reduction.

\paratitle{DRAM Frequency and Voltage Scaling} Many prior works propose to
reduce DRAM energy by adjusting the memory channel frequency and/or the DRAM
supply voltage dynamically. Deng et al.~\cite{deng-asplos2011} propose MemScale,
which scales the frequency of DRAM at runtime based on a performance predictor
of an in-order processor. Other work focuses on developing management policies
to improve system energy efficiency by coordinating DRAM \emph{DFS} with DVFS on
the CPU~\cite{deng-micro2012, deng-islped2012, begum-iiswc2015} or
GPU~\cite{paul-isca2015}. In addition to frequency scaling, David et
al.~\cite{david-icac2011} propose to scale the DRAM supply voltage along with
the memory channel frequency, based on the memory bandwidth utilization of
applications.

In contrast to all these works, our work focuses on a detailed experimental
characterization of real DRAM chips as the supply voltage varies. Our study
provides fundamental observations for potential mechanisms that can mitigate
DRAM and system energy consumption. Furthermore, frequency scaling hurts memory
throughput, and thus significantly degrades \fixIV{the} system performance of
\fixIV{especially} memory-intensive workloads (see \new{Section 2.4 in our
  SIGMETRICS 2017 paper~\cite{chang-sigmetrics2017}} for
our quantitative analysis). We demonstrate the importance \fix{and benefits}
of \fix{exploiting} our \fix{experimental} observations by proposing \voltron,
one \fixIV{new example} mechanism that uses our observations to reduce DRAM and
system energy without sacrificing memory throughput.

\paratitle{Low-Power Modes for DRAM} Modern DRAM chips support various low-power
standby modes. \fix{Entering and exiting these modes incurs some amount of
  latency, which delays memory requests that must be serviced.} To increase the
opportunities to exploit these low-power modes, several prior works propose
mechanisms that increase periods of memory idleness through data placement
(e.g.,~\cite{lebeck-asplos2000, fan-islped2001}) and memory traffic reshaping
(e.g.,~\cite{aggarwal-hpca2008, bi-hpca2010, amin-islped2010, lyuh-dac2004,
  diniz-isca2007}). Exploiting low-power modes is orthogonal to our work on
studying the impact of \fixIV{reduced-voltage operation in DRAM}. Furthermore,
low-power modes have a smaller effect on memory-intensive workloads, which
exhibit little idleness in memory accesses, \fixIV{whereas, as we \new{show in
  Section~\ref{sec:voltron:eval}}, our mechanism is especially effective on memory-intensive
  workloads.}

\paratitle{Low-Power DDR DRAM Chips} Low-power DDR (LPDDR)~\cite{jedec-lpddr3,
jedec-lpddr4, patel-isca2017} is a specific type of DRAM that is optimized for
low-power systems like mobile devices. To reduce power consumption, LPDDRx
(currently in its 4th generation) employs a few major design changes that differ
from conventional DDRx chips. First, LPDDRx uses a low-voltage swing I/O
interface that consumes 40\% less I/O power than DDR4
DRAM~\cite{choi-memcon2013}. Second, it supports additional low-power modes with
a lower supply voltage. Since the LPDDRx array design remains the same as DDRx,
our observations on the correlation between access latency and array voltage are
applicable to LPDDRx DRAM \fixIV{as well}. \voltron, our proposal, can provide
significant benefits in LPDDRx, \fixIII{since array} energy consumption is
significantly \emph{higher} than the energy consumption of \fixIII{peripheral
  circuitry in LPDDRx chips}~\cite{choi-memcon2013}. \changes{We leave the
  detailed evaluation of LPDDRx chips for future work since our current
  experimental platform is not capable of evaluating them. \new{\ch{Two recent
    experimental works~\cite{patel-isca2017, kim-hpca2018} examine} the retention time behavior
    of LPDDRx chips and find it to be similar to DDRx chips.}}

\paratitle{Low-Power DRAM Architectures} \fixIV{Prior} works
(e.g.,~\cite{udipi-isca2010, zhang-isca2014, cooper-balis-ieeemicro2010,
  chatterjee-hpca2017}) propose to modify the DRAM chip architecture to reduce
the \act power by activating only a fraction of a row instead of the entire row.
Another common technique, called sub-ranking or \new{mini-ranks}, reduces dynamic DRAM power by
accessing data from a subset of chips from a DRAM module~\cite{zheng-micro2008,
  yoon-isca2011, ware-iccd2006}. \fix{A couple of} prior
works~\cite{malladi-isca2012, yoon-isca2012} propose DRAM module architectures
that integrate many low-frequency LPDDR chips to enable DRAM power reduction.
These proposed changes to DRAM chips or DIMMs are orthogonal to our work.

\fix{ \paratitle{Reducing Refresh Power} In modern DRAM chips, although
different DRAM cells have widely different retention times~\cite{liu-isca2013,
  kim-edl2009, patel-isca2017}, memory controllers conservatively refresh
\emph{all} of the cells based on the retention time of a small fraction of weak
cells, which have the longest retention time out of all of the cells. To reduce
DRAM refresh power, many prior works (e.g., \new{\cite{liu-isca2012,
    agrawal-hpca2014, qureshi-dsn2015, liu-isca2013, baek-tc2014,
    khan-micro2017, mutlu-date2017, superfri,
    venkatesan-hpca2006,bhati-isca2015,lin-iccd2012,ohsawa-islped1998,
    patel-isca2017, khan-sigmetrics2014, khan-dsn2016, khan-cal2016}}) propose
mechanisms to reduce unnecessary refresh operations, and, thus, refresh power,
by characterizing the retention time profile (i.e., the \fixIV{data} retention
behavior of each cell) within the DRAM chips. However, these techniques do not
reduce the power of \emph{other} DRAM operations, and these prior works do
\emph{not} provide an experimental characterization of the effect of reduced
voltage level\fixIV{s} on \fixIV{data} retention time.


\paratitle{Improving DRAM Energy Efficiency by Reducing Latency or Improving
Parallelism} \fixIV{Various} prior works (e.g.,\new{~\cite{hassan-hpca2017,
  lee-hpca2015, chang-hpca2014, lee-thesis2016, lee-hpca2013, hassan-hpca2016,
  seshadri-micro2017, seshadri-micro2015, mutlu-imw2013,
  lee-sigmetrics2017,kim-isca2012, lee-taco2016,
  seshadri-micro2013, chang-hpca2016, lee-pact2015}}) improve DRAM energy
  efficiency by reducing the execution time through techniques that reduce the
  DRAM access latency or improve parallelism between memory requests.  These
  mechanisms are orthogonal to ours, because they do \fixIV{not reduce} the
  voltage level of DRAM. }

\new{

  \paratitle{Improving Energy Efficiency by Processing in Memory} Various prior
works~\cite{ahn-isca2015,ahn-isca2015-2,7056040,7429299,guo-wondp14,592312,
  seshadri-cal2015,mai-isca2000,draper-ics2002,
  seshadri-micro2015,hsieh-iccd2016,hsieh-isca2016,
  amirali-cal2016, stone-1970, fraguela-2003,375174,808425,
  4115697,694774,sura-2015,zhang-2014,akin-isca2015,
  babarinsa-2015,7446059,6844483,pattnaik-pact2016, seshadri-thesis2016,
  seshadri-micro2017, chang-hpca2016,kim-apbc2018,hashemi-isca2016, ami-asplos2018} examine
processing in memory to improve energy efficiency. Our analyses and techniques
can be combined with these works to enable low-voltage operation in
processing-in-memory engines.

}

\fixIII{\paratitle{Experimental Studies of DRAM Chips} Recent works
\fixIV{experimentally investigate various} reliability, \fixIV{data} retention,
and latency characteristics of modern DRAM chips~\cite{liu-isca2012,
  liu-isca2013, kim-isca2014, chang-sigmetrics2016, lee-hpca2015,
  lee-sigmetrics2017, chandrasekar-date2014, khan-sigmetrics2014,
  jung-memsys2016, jung-patmos2016,
  hassan-hpca2017,khan-dsn2016,patel-isca2017,lee-thesis2016,kim-thesis,
  meza-dsn2015, schroeder-sigmetrics2009, sridharan-asplos2015,
  sridharan-sc2012} \new{usually using FPGA-based DRAM testing infrastructures,
like SoftMC~\cite{hassan-hpca2017}, or using large-scale data from the field}.
None of these works study these characteristics under reduced-voltage operation,
which we do in this paper.}


\fixIII{\paratitle{Reduced-Voltage Operation in SRAM Caches} Prior works propose
different techniques to enable SRAM caches to operate under reduced voltage
levels
(e.g.,~\cite{alameldeen-isca2011,alameldeen-tc2011,wilkerson-ieeemicro2009,chishti-micro2009,roberts-dsd2007,wilkerson-isca2008}).
These works are orthogonal to our experimental study because we focus on
\fixIV{understanding and enabling reduced-voltage operation in} DRAM, which is a
\fixIV{significantly} different memory technology than SRAM.}

\section{Significance}

\new{Our SIGMETRICS 2017 paper~\cite{chang-sigmetrics2017} presents a new set of
detailed experimental characterization and \ch{analyses} on the
voltage-latency-reliability trade-offs in modern DRAM chips. In this section, we
describe the potential impact that our study can bring to the research community
and industry.}

\subsection{Potential Industry Impact}

We believe our experimental characterization results and proposed mechanism can
have significant impact in fast-growing data centers \new{as well as mobile
  systems}, where DRAM power consumption is growing due to higher demand for
memory capacity for certain types of service (e.g., memcached). To reduce the
energy and power consumed by DRAM, DRAM manufacturers have been decreasing the
supply voltage of DRAM chips with newer DRAM standards (e.g., DDR4) or
low-voltage variants of DDR, such as LPDDR4 (Low-Power DDR4) and DDR3L (DDR3
Low-voltage). However, the supply voltage reduction has been conservative with
each new DDR standard, which takes years to be adopted by the vendors and the
market. For example, since the release of DDR3L (1.35V) in 2010, the supply
voltage has reduced by \emph{only} 11\% with the latest DDR4 standard (1.2V)
released in 2014. Furthermore, since the release of DDR4 in 2014, the supply
voltage for most commodity DDR4 chips has remained at 1.2V. As a result, further
reducing DRAM supply voltage below the standard voltage, \new{as we do in our
  SIGMETRICS 2017 paper~\cite{chang-sigmetrics2017}}, can be a \new{very} effective way
of reducing DRAM power consumption. However, to do so, we need to \new{carefully
and rigorously} understand how DRAM
chips behave under reduced-voltage operation.

To enable the development of new mechanisms that leverage reduce-voltage
operation in DRAM, we provide \new{the first set of} comprehensive experimental
results on the effect of using a wide range of different supply voltage values
on the reliability, latency, and retention characteristics of DRAM chips. In
this work, we demonstrate how we can use our experimental data to design a new
mechanism, Voltron (\secref{voltron}), which reduces DRAM energy consumption
through voltage reduction. Therefore, we believe that understanding and
leveraging reduced-voltage operation will help industry improve the energy
efficiency of memory subsystems.

\subsection{Potential Research Impact}

Our paper sheds \new{new} light on the feasibility of enabling reduced-voltage
operation in manufactured DRAM chips. One important research question that our
work raises is \textit{how do modern DRAM chips behave under a wide range of
  supply voltage levels?} Existing systems are limited to a few DRAM power
states, which prevent DRAM from serving memory accesses when it enters a
low-power state. However, in our work, we show that it is possible to operate
commodity DRAM chips under a wide range of supply voltage levels while still
being able to serve memory accesses under a different set of trade-offs. To
facilitate further research initiative to exploit reduced-voltage operation in
DRAM chips, we have open-sourced our characterization results, FPGA-based
testing platform~\cite{hassan-hpca2017}, and DRAM SPICE circuit model (for
validation) in our GitHub repository~\cite{safari-github}. We believe that these
tools can be extended for other research objectives besides studying voltage
reduction in DRAM. \new{One potential direction is to leverage our results to
  design mechanisms that reduce DRAM latency by operating DRAM at a higher supply
  voltage. }

\subsection{Applicability to Other Memory Technologies}

\new{We believe the high-level ideas of our work can be \ch{leveraged} in the context
of other memory technologies, such as NAND flash memory~\cite{cai-ieee2017, cai-ieeearxiv2017, cai-bookchapter2017},
PCM\ch{~\cite{lee-isca2009, lee-ieeemicro2010, qureshi-isca2009,
  yoon-taco2014,lee-cacm2010,qureshi-micro2009,yoon-iccd2012, meza-weed2013}},
\mbox{STT-MRAM}\ch{~\cite{ku-ispass2013,guo-isca2009,chang-hpca2013, meza-weed2013,
naeimi-itj2013}},
RRAM~\cite{wong-ieee2012}, \ch{or hybrid memory
systems~\cite{meza-weed2013, qureshi-isca2009, yoon-iccd2012, ramos-ics2011,
zhang-pact2009, li-cluster2017, yu-micro2017, jiang-hpca2010, phadke-date2011,
agarwal-asplos2017, dulloor-eurosys2016, pena-cluster2014, bock-iccd2016,
gai-hpcc2016, liu-iccd2016}}. A recent work on NAND flash memory, for example,
proposes reducing the pass-through voltage~\cite{cai-dsn2015,cai-ieee2017, cai-ieeearxiv2017, cai-bookchapter2017} to
reduce read disturb errors, which in turn saves energy. We refer the reader to
past works on NAND flash memory for a more detailed analysis of
reliability-voltage
trade-offs~\cite{cai-dsn2015,cai-hpca2015,cai-hpca2017,cai-ieee2017, cai-ieeearxiv2017, cai-bookchapter2017}. We hope
our work inspires characterization and understanding of reduced-voltage
operation in other memory technologies, with the goal of enabling a more
energy-efficient system design.}

\ignore{Another key impact of our paper is that we demonstrate the effectiveness
of utilizing real-world characterization to design new optimization techniques
for memory subsystems. Given that a diverse set of new memory technologies
(e.g., STT-RAM) are becoming promising substrates for disrupting the once
clear-cut memory hierarchy, we believe that characterizing these new memory
technologies is a new and important research direction.

We hope that the experimental characterization, analysis, and optimization
techniques presented in this paper will enable the development of other new
mechanisms that can effectively exploit the trade-offs between voltage,
reliability, and latency in DRAM to improve system performance, efficiency,
and/or reliability.
}


\section{Conclusion}

\changes{ \new{Our SIGMETRICS 2017} paper~\cite{chang-sigmetrics2017} provides the
first experimental study that comprehensively characterizes and analyzes the
behavior of DRAM chips when the supply voltage is reduced below its nominal
value. We demonstrate, \fixIV{using 124 DDR3L DRAM chips}, that the \fix{DRAM}
supply voltage can be reliably reduced to a certain level, beyond which errors
arise within the data. We then experimentally demonstrate the relationship
between the supply voltage and the latency of the fundamental DRAM operations
(activation, restoration, and precharge). \fixIV{We show that bit errors caused
  by reduced-voltage operation can be eliminated by increasing the latency of
  the three fundamental DRAM operations.} By \fix{changing} the memory
controller configuration to allow for the longer latency of these operations, we
can thus \emph{further} lower the supply voltage without inducing errors in the
data. We also \fixIV{experimentally} characterize the relationship between
reduced supply voltage and error locations, stored data patterns, temperature,
and data retention.


Based on these observations, we propose and evaluate \voltron, a low-cost energy
reduction mechanism that reduces DRAM energy \emph{without} \fix{affecting}
memory data throughput. \voltron reduces the supply voltage for \emph{only} the
DRAM array, while maintaining the nominal voltage for the peripheral circuitry
to continue operating the memory channel at a high frequency. \voltron uses a
new piecewise linear performance model to find the array supply voltage that
maximizes the system energy reduction within a given performance loss target.
\fix{Our experimental evaluations across a wide variety of workloads}
demonstrate that \voltron significantly reduces system energy \fix{consumption}
with only \fix{very} modest performance loss.

We conclude that it is very promising to understand and exploit reduced-voltage
operation in modern DRAM chips. We hope that the experimental characterization,
analysis, and optimization techniques presented in \new{our SIGMETRICS 2017}
paper will enable the development of other new mechanisms that can effectively
\fixIII{exploit the trade-offs between voltage, reliability, and latency in DRAM
  to improve system performance, efficiency, and/or reliability. \new{We also
    hope that our paper's studies inspire new experimental studies to understand
    reduced-voltage operation in other memory technologies, such as NAND flash
    memory, PCM, and STT-MRAM.}}}

\section*{Acknowledgments}

We thank \ch{the anonymous reviewers of SIGMETRICS
2017 and SAFARI group members for their} feedback. We acknowledge the support of
Google, Intel, \fix{NVIDIA}, Samsung, VMware, and the United States Department
of Energy. This research was supported in part by the ISTC-CC, SRC, and NSF
(grants 1212962 and 1320531). Kevin Chang was supported in part by an SRCEA/Intel
Fellowship.

\balance 
{
\bstctlcite{bstctl:etal, bstctl:nodash, bstctl:simpurl}
\bibliographystyle{IEEEtranS}
\bibliography{kevin_paper}
}

\end{document}